\documentclass[12pt,preprint]{aastex}

\newcommand{\hmol}{H{${_2}$}}		
\newcommand{\carone}{C~{\footnotesize{I}}}  	
\newcommand{\cartwo}{C~{\footnotesize{II}}}  	
\newcommand{\carthree}{C~{\footnotesize{III}}}  
\newcommand{\nitone}{N~{\footnotesize{I}}}	
\newcommand{\nittwo}{N~{\footnotesize{II}}}	
\newcommand{\nitthree}{N~{\footnotesize{III}}}	
\newcommand{\oxyone}{O~{\footnotesize{I}}}  	
\newcommand{\oxytwo}{O~{\footnotesize{II}}}  	
\newcommand{\oxythree}{O~{\footnotesize{III}}}  
\newcommand{\oxyfour}{O~{\footnotesize{IV}}}  
\newcommand{\oxyfive}{O~{\footnotesize{V}}}  	
\newcommand{\oxysix}{O~{\footnotesize{VI}}}  	
\newcommand{\oxyseven}{O~{\footnotesize{VII}}}  
\newcommand{\magtwo}{Mg~{\footnotesize{II}}}    
\newcommand{\sitwo}{Si~{\footnotesize{II}}}	
\newcommand{\stwo}{S~{\footnotesize{II}}}	
\newcommand{\sthree}{S~{\footnotesize{III}}}	
\newcommand{\sfour}{S~{\footnotesize{IV}}}	
\newcommand{\ssix}{S~{\footnotesize{VI}}}	
\newcommand{\irontwo}{Fe~{\footnotesize{II}}}   
\newcommand{\ironthree}{Fe~{\footnotesize{III}}}   
\newcommand{\fuse}{{\it{FUSE}}}			

\shorttitle{OVI LB}
\shortauthors{Shelton}
\begin{document}

\title{Surprisingly Little  \oxysix\  Emission 
Arises in the Local Bubble}

\author{R. L. Shelton\altaffilmark{1}
\affil{Department of Physics and Astronomy, Johns Hopkins University,
	3400 North Charles Street, Baltimore, MD 21218}
\email{shelton@pha.jhu.edu}}

\begin{abstract}

This paper reports the first study of the \oxysix\ resonance line
emission ($\lambda\lambda$ 1032, 1038) originating in
the Local Bubble (or Local Hot Bubble) surrounding the solar neighborhood.
In spite of the fact that \oxysix\ absorption within the Local Bubble
has been observed,
no resonance line emission was detected during
our 230 ksec {\it{Far Ultraviolet Spectroscopic Explorer}}
observation of a ``shadowing'' filament in the southern Galactic
hemisphere.  As a result, tight 2 sigma upper limits are set on 
the intensities in the 1032 and 1038 \AA\ emission lines:
500 and 530 photons cm$^{-2}$ s$^{-1}$ sr$^{-1}$, respectively.
These values place strict constraints on models and simulations.  They
suggest that the \oxysix-bearing plasma and the
X-ray emissive plasma reside in distinct regions of the Local
Bubble and are not mixed in a single plasma, whether in
equilibrium with $T \sim10^6$~K or highly overionized with 
$T \sim 4$ to $6 \times 10^4$~K.
If the line of sight intersects multiple cool clouds within the
Local Bubble, then the results also suggest that the hot/cool
transition zones differ from those in current simulations.
With these intensity upper limits, we establish limits on the
electron density, thermal pressure, pathlength, and cooling
timescale of the \oxysix-bearing plasma in the Local Bubble.
Furthermore, the intensity of \oxysix\ resonance line doublet photons
originating in the Galactic thick disk and halo is determined
(3500 to 4300 photons cm$^{-2}$ s$^{-1}$ sr$^{-1}$),
and the electron density, thermal pressure, pathlength, and cooling
timescale of its \oxysix-bearing plasma are calculated.
The pressure in the Galactic halo's \oxysix-bearing plasma
(3100 to 3800 K cm$^{-3}$)
agrees with model predictions for the {\it{total}} pressure
in the thick disk/lower halo.
We also report the
results of searches for the emission 
signatures of interstellar
\carone, \cartwo, \carthree,
\nitone, \nittwo, \nitthree, \magtwo, \sitwo,
\stwo, \sthree, \sfour, \ssix, \irontwo, and \ironthree.

\end{abstract}




\section{Introduction}

\begin{figure}
\plotone{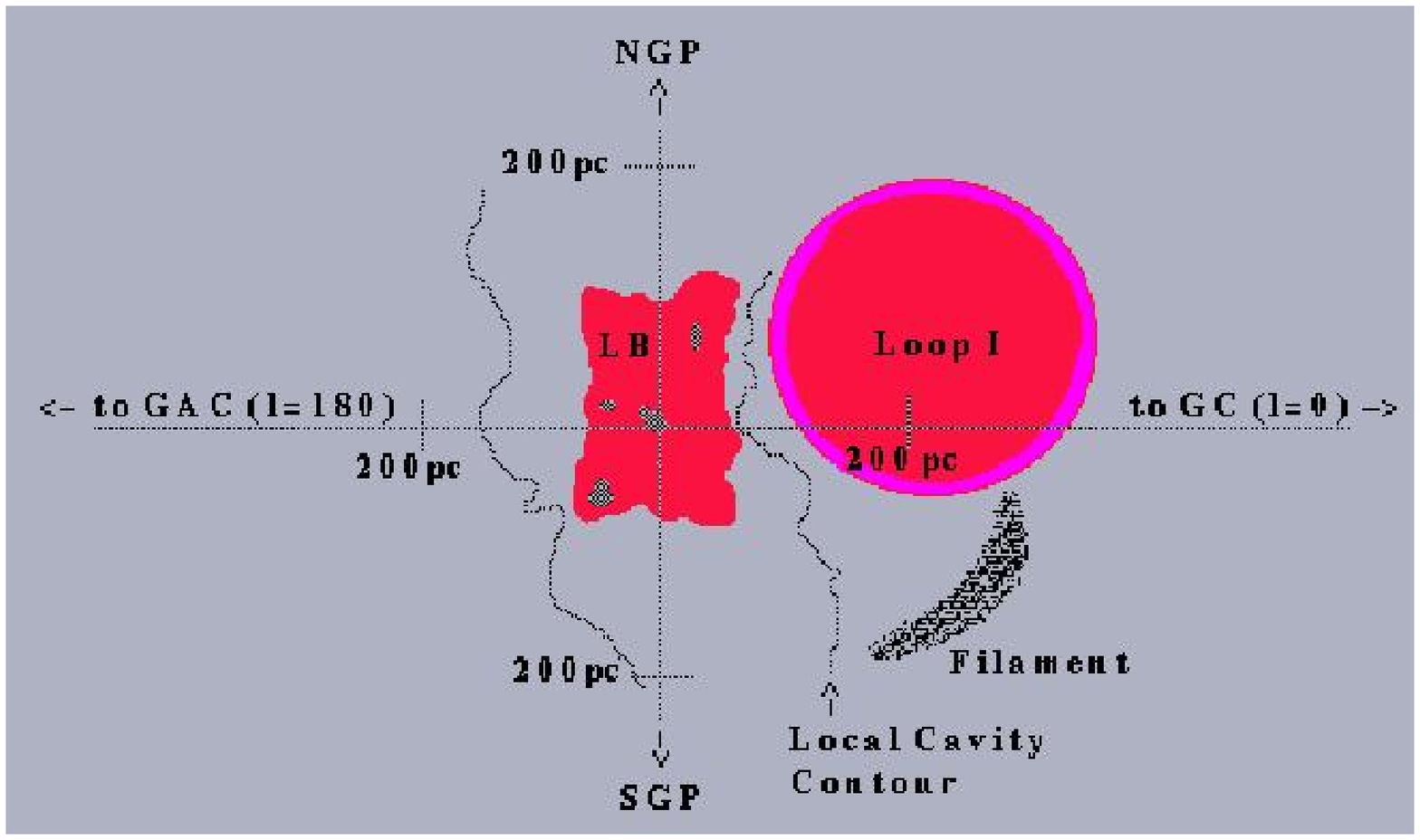}
  \caption{The Local Bubble (LB) is sketched in red in the center of the
cartoon.
The LB's size and shape were drawn from 
\citet{snowden_etal_98}.
The gray spots within the LB represent 
the cool clouds.
The dotted, wiggly lines beyond the Local Bubble depict the
outline of the Local Cavity, whose shape and size has been taken from the
50 m\AA\ Na I D2 projected contours in \citet{sfeir_etal}.
To the right and lower right of the LB are cartoons representing 
the nearby Loop I superubble and the observed filament.
Note that the filament is well beyond the calculated extent of
the Local Bubble.
An axis has been superimposed.  
The Sun lies at the origin.  The Galactic center (GC) is to the
right, the Galactic anticenter (GAC) is to the left, the
north Galactic pole (NGP) is at the top, and the South Galactic pole (SGP)
is at the bottom.  Each of the tick marks is 200 pc from the center of the
grid.    }
\label{fig:schematic}
\end{figure}

The Local Bubble (LB) or Local Hot Bubble and its environment are
sketched in Figure~\ref{fig:schematic}.
The Local Bubble, a $\sim10^6$~pc$^{3}$
region of X-ray emissive, presumably hot ($\sim10^6$~K)
plasma, is nestled within a rarefied cavity in the Galactic disk 
called the Local Cavity, 
\citep{knapp,mccammon_etal,snowden_etal_98,sfeir_etal}.  
Within the Local Bubble, lie
a number of cool ($\sim10^4$~K) clouds,
including the complex of parsec-scale clouds surrounding
the Sun \citep{frisch_86,lallement_bertin,gry_etal}. 
Estimates of the number of clouds per line of sight range from
$\sim2$ to $\sim6$, with $\sim2$ coming from the
ratio of the typical observed absorbing column density 
\citep{hutchinson_etal}
to the absorbing column density of the local cloud
\citep{lallement_etal},
and $\sim6$ coming from the number of clouds along
the $\epsilon$ CMa line of sight \citep{gry_etal}.
Beyond the Local Bubble, in the direction of the Galactic center, 
lies a superbubble named Loop I, which 
was blown by the stars and supernovae
in the Sco Cen association \citep{egger}.

The Local Bubble 
is larger and more energetic than a supernova remnant, yet smaller
and less energetic than a superbubble.
Such bubbles are important as
most of the hot gas in the Galactic disk resides within them.
Like the distribution of small versus large stellar associations, 
the population of bubbles is likely to be larger than
the population of superbubbles.  
Nonetheless,
because its softer
X-ray photons from its relatively cooler plasma
are more easily absorbed, a bubble 
like the Local Bubble
would be much more difficult to detect from a great distance than would
an energetic superbubble like Loop I.  
Thus, many other Local Bubble analogs may reside within the
Galaxy, unknown to us because they are obscured by the 
neutral and molecular material of the Galactic disk.  
In this sense,
we are fortunate to have a local specimen to examine.

At this time, our understanding of the Local Bubble is chiefly
phenomenological.
To some degree, the Local Bubble is defined as the source of
soft X-rays ($\sim \frac{1}{4}$~keV) produced between us and the nearest 
opaque material.
Its physical presence has been surmised from 
the anti-correlation between soft X-ray intensities 
and 
neutral hydrogen column densities (associated with the larger Local Cavity),
detections of X-rays originating in the foreground of opaque clouds,
relative constancies between 
Be (0.07 to 0.11 keV), B (0.13 to 0.19 keV) and C (0.16 to 0.28 keV)
band surface brightnesses across the sky (implying
very little absorption as the effective absorption cross sections vary
significantly between the bands), 
and consistency among two decades of X-ray observations.

Fundamental characteristics of the LB, 
such as its
presumed temperature and size, cannot be determined directly, but can only be
estimated when additional constraints (such as ionizational equilibrium, 
constancy of emissivity, 
and filling factor) are assumed.
Given these assumptions, the estimated temperature has been derived from the
ratio of fluxes in broad soft X-ray bands.  
In directions where the Galactic disk
blocks all of the $\frac{1}{4}$~keV emission from beyond the Local Bubble,
the observed {\it{ROSAT}} R2 to R1 band ratio is
$\sim 1.06$, implying a temperature of $10^{6.1}$~K \citep{kuntz_snowden}.  
This temperature
is consistent with values found by shadowing observations (described
below) at higher Galactic latitudes \citep{kuntz_snowden,snowden_etal_00}
and with Wisconsin's {\it{Ultrasoft X-ray Telescope}} ({\it{UXT}})
and All-Sky Survey's 
B/Be and C/B band ratios
\citep{bloch_etal,juda_etal}.
Similar temperatures were also derived from the
{\it{Diffuse X-ray Spectrometer}}'s 
soft X-ray spectra, although these high
resolution spectra were difficult to fit with simple models
\citep{sanders_etal}.


Desirous of a better understanding of our environs as well as
a model of the bubble population (useful for calculating the hot
gas filling factor),
both observers and theorists have long sought explanations of
the LB's origin and life history.  Many researchers have pointed 
out that 
the energy embodied in the Local Bubble's X-ray emitting plasma
is so great as to require a supernova or a series of supernovae, possibly aided
by stellar winds.  The energy may have been deposited within an existing gap 
in the 
H I disk or may have pushed aside the disk material to create a cavity.  
Alternatively, the explosion(s) and winds may have originated in the 
nearby Loop I (Sco-Cen) superbubble
and blown into the solar region \citep{frisch_81,bochkarev}, or
the Local Bubble and Loop I may have been
blown independently but are now interacting \citep{egger_aschenbach}.
In some of these scenarios, 
the X-ray emitting plasma may have expanded after being shock heated, 
vastly cooling the plasma, retarding the
collisional recombination rates, and thus ``freezing'' the plasma into an
``overionized'' state \citep{breitschwerdt_schmutzler,breitschwerdt}.  
In contrast, if the expansion was restrained and the 
time-scale was long, the plasma may have already approached collisional 
ionizational equilibrium \citep{smith_cox}.  
These two cases should produce markedly different 
spectral signatures.
Presently, many theorists and observers are working to 
decipher the Local Bubble's life story by comparing its X-ray 
spectra to spectra predicted by hydrodynamical simulations.
It is hoped that
when the observed fluxes in multiple bands or emission lines are matched by
the predictions of a particular type of model, then we will have found a key to
deciphering the story of the Local Bubble.
Here, we extend this sort of comparison into the ultraviolet, using the 
\oxysix\ resonance line (1032, 1038 \AA) intensities.
We also compare the intensities with \oxysix\ absorption line column densities
taken from the literature, in order to estimate the
physical properties of the \oxysix-bearing plasma.


We use the ``shadowing'' strategy long employed in X-ray
analyses to isolate
the Local Bubble's emission from the more distant
interstellar medium's emission (see Section 2).
The observations (228 ksec toward $l = 278.6^{\rm{o}},b = -45.3^{\rm{o}}$)
and data reduction techniques are discussed in Section 3.
Although we observed for a very long time, no \oxysix\ resonance
line emission was detected.  Tight upper limits 
were placed on the 1032 and 1038 \AA\ lines
(see Section 4).  The results of our searches for other cosmic lines within
the bandpass are also reported (see Section 4).
Our upper limit on the \oxysix\ doublet intensity is well below
the $4 \times 10^4$~K, overionized model \citep{breitschwerdt}
predictions.  This discrepancy, as well as discrepancies in the
\carthree\ intensities and the \oxysix\ column density, 
practically disallow models of this sort.
Our two sigma upper limit on the \oxysix\ doublet intensity is only
marginally above the minimum intensity expected from the 
\oxysix-rich zone around the most local cool cloud 
\citep{slavin} combined with the \oxysix-rich regions 
in \citet{smith_cox}'s suite of simulations of
a multiple supernovae induced hot bubble.
In the event that the line of sight intersects multiple cool clouds
within the Local Bubble, as well as the Local Bubble's wall, then
our doublet upper limit is well below the predicted intensity
and places a strong constraint on models of hot gas/cool gas transition
zones.
In order to determine if the upper limits require the Local Bubble
to be undergoing unusual physical processes or contain unusual
plasmas, we constructed and compared with a generalized
multi-component 
(an \oxysix-rich $\sim 3\times 10^5$~K plasma and a soft X-ray 
emitting $\sim 10^6$~K plasma) 
model of the Local Bubble.  This model's \oxysix\ intensity and column density
and soft X-ray surface brightness were all consistent with observations
(see Section 5).
By combining our \oxysix\ intensity upper limit 
with other researcher's \oxysix\ column density estimates,
we were able to calculate 
limits on the electron density, thermal pressure, pathlength,
and cooling timescale
(also see Section 5).
By subtracting the Local Bubble's \oxysix\ doublet intensity upper 
limit from measurements for longer, high latitude lines of sights, we
found the halo's intensity, from which its electron density,
thermal pressure, pathlength, and cooling rate 
were estimated.  
The results of this project are summarized in Section 6.


\section{Shadowing Strategy}

\begin{figure}
\epsscale{0.45}
\plotone{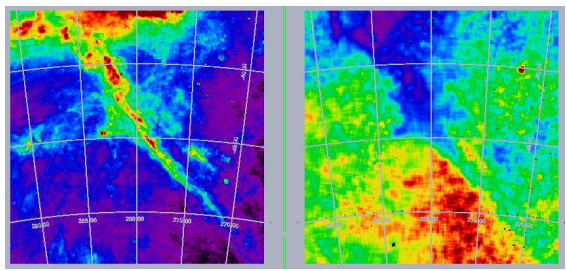}
  \caption{Right Panel:
%
Map of the {\it{DIRBE}}-corrected {\it{IRAS}} 100 $\mu$m data for 
a portion of the sky centered on $l \sim 280^{\rm{o}}$, 
$b \sim -45^{\rm{o}}$. 
The ``shadowing'' filament discussed in this paper 
runs diagonally across the image.
Left Panel:  Map of the {\it{ROSAT}} $\frac{1}{4}$~keV data 
for the same projection.
Regions of missing data are black.
By comparing these two maps,
we can see that the ``shadowing'' filament corresponds with the
sharply delineated arch of low $\frac{1}{4}$~keV surface brightness in
the soft X-ray map where
the filament blocks soft X-ray emission from the
Galactic halo.  The finite ``zero-point'' surface brightness
($\sim 400 \times 10^{-6}$ counts s$^{-1}$ arcmin$^{-2}$) is 
due to the Local Bubble emission. 
Both panels are adaptations of figures in \citet{snowden_etal_97}.
}
\label{fig:diras}
\end{figure}

	The intensity of resonance
line photons from \oxysix\ along extended lines of sight
through the Local Bubble, the Galactic halo, 
and intervening regions is reported elsewhere 
\citep{shelton_etal,dixon_etal,shelton,welsh_etal}.
Here, we wish to measure {\bf{only}} the \oxysix\ intensity of the Local 
Bubble, and so must block the photons from external regions.   
To do so, we employ the ``shadowing'' strategy
oft-used in X-ray analyses.  In the ideal shadowing observation,
an opaque object blocks the emission originating beyond it,
so that the observed emission originates between the viewer and the object.
For the shadowing object, we have chosen a filament in the 
Southern Galactic hemisphere
(see Figure~\ref{fig:diras}).
\citet{penprase_etal} estimated the distance to the filament as
230$\pm$30 pc, which places the filament between the Local Bubble and
the Galactic halo.  
\citet{penprase_etal} also estimated the mean color excess 
($E(B-V)$) of the filament as
0.17$\pm$0.05 magnitudes, implying an absorbing column density
of $N_H = 8.4 \pm 2.5 \times 10^{20}$~cm$^{-2}$.  This is
consistent with the column density implied by the \citet{schlegel_etal}
{\it{IRAS}} 100 $\mu$m measurement of
$\sim 7.3$ MJy sr$^{-1}$, which when scaled by the
100 $\mu$m to $N_H$ converstion relation for the southern
hemisphere listed in \citet{snowden_etal_00}, yields a column 
density of $9.9 \times 10^{20}$ cm$^{-2}$. 
For this color excess and the extinction curve and parameterization 
presented in \citet{fitzpatrick}, 
$89^{+5}_{-11}\%$ of the \oxysix\ photons originating beyond the
filament should be blocked.

As a demonstration of the filament's ``shadowing'' function, 
we note it's ability to block soft X-rays.
(The theoretical extinction of $\sim\frac{1}{4}$ X-rays is
larger than that of ultraviolet photons.)
The {\it{ROSAT}} $\frac{1}{4}$~keV band 
image (see Figure~\ref{fig:diras}) reveals 
an obvious soft X-ray depression coincident with
the filament.  The surface brightness seen in the direction
of the filament and attributed to the Local Bubble
is $590 \times 10^{-6}$~counts s$^{-1}$ arcmin$^{-2}$,
while the ``off-filament'' surface brightness 
is $\sim 1300 \times 10^{-6}$~counts s$^{-1}$ arcmin$^{-2}$
(K. D. Kuntz, private communication).
Not surprisingly, the filament has already been used 
successfully in {\it{ROSAT}} soft X-ray shadowing analyses 
\citep{wang_yu,snowden_etal_00}.

One potential difficulty in using this region of the sky is posed by
\citet{penprase_etal}'s suggestion that the
filament may be part of a supernova remnant.  If this were the
case, then 
the foreground \oxysix\ intensity should be attributed to both the
Local Bubble and the supposed supernova remnant.
Because 
the results or our study are upper limits, this point is unimportant.
Observations of the targeted region, the bulbous portion of the 
filament around $l = 278.6^{\rm{o}},b = -45.3^{\rm{o}}$, are discussed in
the following section.

\section{Observations and Data Reduction}

The \fuse\ focal plane assembly contains four spectrograph
entrance apertures.  The largest, the low resolution (LWRS) aperture
measuring 30'' $\times$ 30'', is used for observations of diffuse
sources.  
Furthermore, the \fuse\ instrument includes multiple approximately co-aligned
channels.  
Four of the channels (LiF 1A, SiC 1A, LiF 2B, and SiC 2B)
cover the 1032, 1038 \AA\ region.  Of these, the LiF 1A is
most sensitive and least prone to scattered light problems
in the 1032, 1038 \AA\ region.
We use this channel for our \oxysix\ analysis.
In addition, we use the LiF 1A, SiC 1B, LiF 2A, and SiC 2A channels
for the \carone, \cartwo, \carthree,
\nitone, \nittwo, \nitthree, \magtwo, \sitwo,
\stwo, \sthree, \sfour, \ssix, \irontwo, and \ironthree\ analyses.

\begin{figure}
\epsscale{1.0}
\plotone{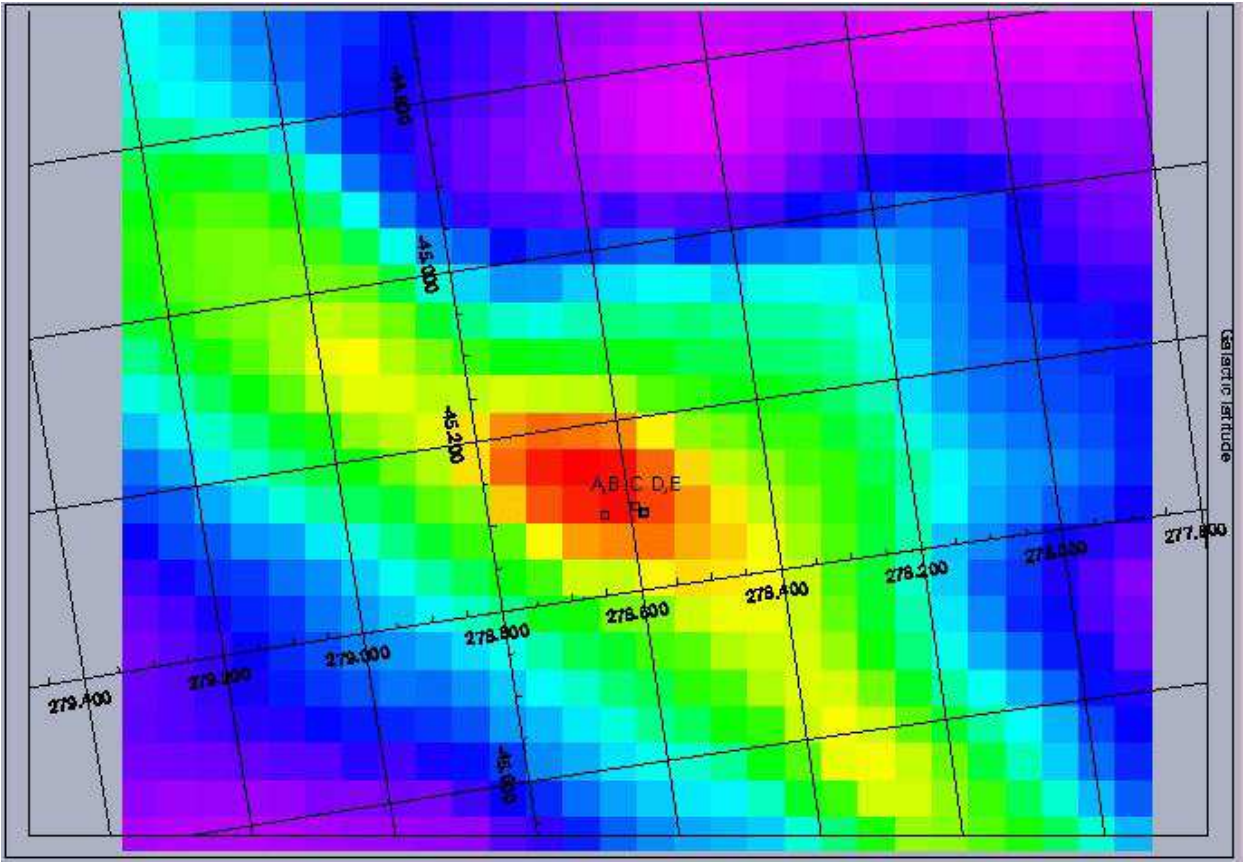} 
  \caption{Infrared 100$\mu$m image of an opaque and bulbous section of the 
filament.
The overlayed boxes indicate the placement of the \fuse\ LWRS aperture.
The B1290101 and B1290102 observations point in the
same direction but with slightly different angles.  
Their aperture location is labeled ``A'' and ``B'' on this image.  
At their location, the 100$\mu$m intensity is 7.2 MJy sr$^{-1}$.
The I2050601 pointing direction is labeled ``C''.  
For this observation, the location of the
\fuse\ aperture crosses a pixel boundary 
in the 100$\mu$m map, such that the average 100$\mu$m intensity is
7.2 MJy sr$^{-1}$.
The I2050510 and I2050501 pointing directions are also nearly identical.
They are labeled ``D'' and ``E'' on the map and have  
100$\mu$m values of 7.0 MJy sr$^{-1}$.
The lowest 100$\mu$m intensity on this image is 2.0
MJy sr$^{-1}$.  Note that the filament's infrared brightness in excess of the
background level implies a degree of obscuration that is 
in very good agreement with \citet{penprase_etal}'s quoted E(B-V).}
\label{fig:pointings}
\end{figure}

The raw data consists of multiple observations of
three nearly coincident directions (see Table~\ref{table:data} and
Figure~\ref{fig:pointings}). 
The combined exposure time (after subtracting time when the
spacecraft was exposed to the Earth's South Atlantic Anomaly,
time when the detectors recorded anomalous bursts, etc.)
for the LiF 1A observations is 228 ksec.  The other detectors 
have had differing useful exposure times.  Most notably,
detector 2 malfunctioned during the I2050501 observation,
reducing the total observation time with the LiF 2A and SiC 2A channels.
The observations can be grouped according to pointing direction.  
Henceforth, the combined I2050501 and I1050510 observations of 
$l = 278.58^{\rm{o}}, b = -45.31^{\rm{o}}$ will be 
denoted by ``I20505'',
the I2050601 observation of 
$l = 278.59^{\rm{o}}, b = -45.30^{\rm{o}}$ will be 
denoted by ``I20506'',
and the combined B1290101 and B1290102 observations of
$l = 278.63^{\rm{o}}, b = -45.31^{\rm{o}}$ will be 
denoted by ``B12901''.
Each of these three datasets was processed using version 2.1 of the CALFUSE
pipeline \citep{sahnow_etal,datahandbook}.
We set the appropriate pipeline switches, so as to 
1.)  exclude counts that produced pulseheights
outside the expected range for cosmic photons (the chosen
CALFUSE pipeline cutoffs were 4 and 12 in the standard
arbitrary units for the older datasets (I20505 and I20506) and
4 to 31 for the newer dataset (B12901)) 
in order to reduce the detector background noise, and
2.)  disallow automatic background subtraction, enabling
us to more accurately calculate the signal to noise of the results.

\begin{figure}
\epsscale{0.45}
\plotone{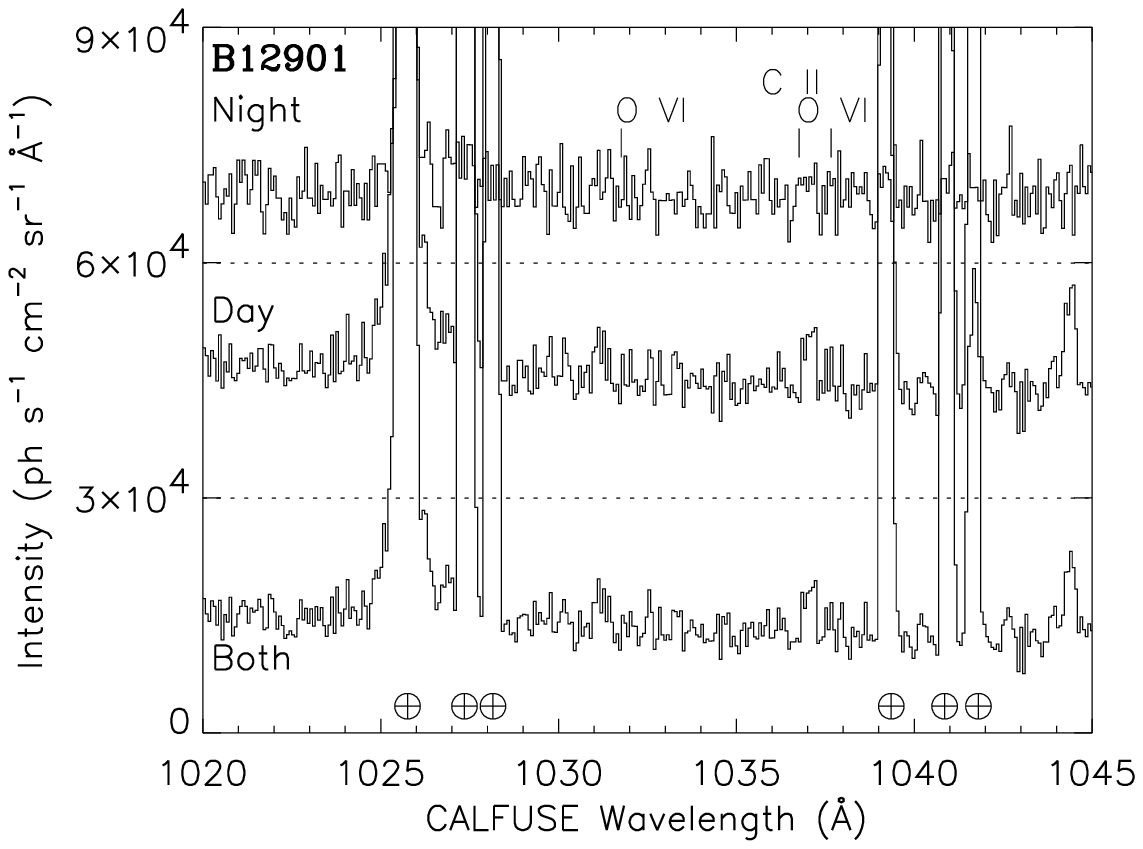}
\plotone{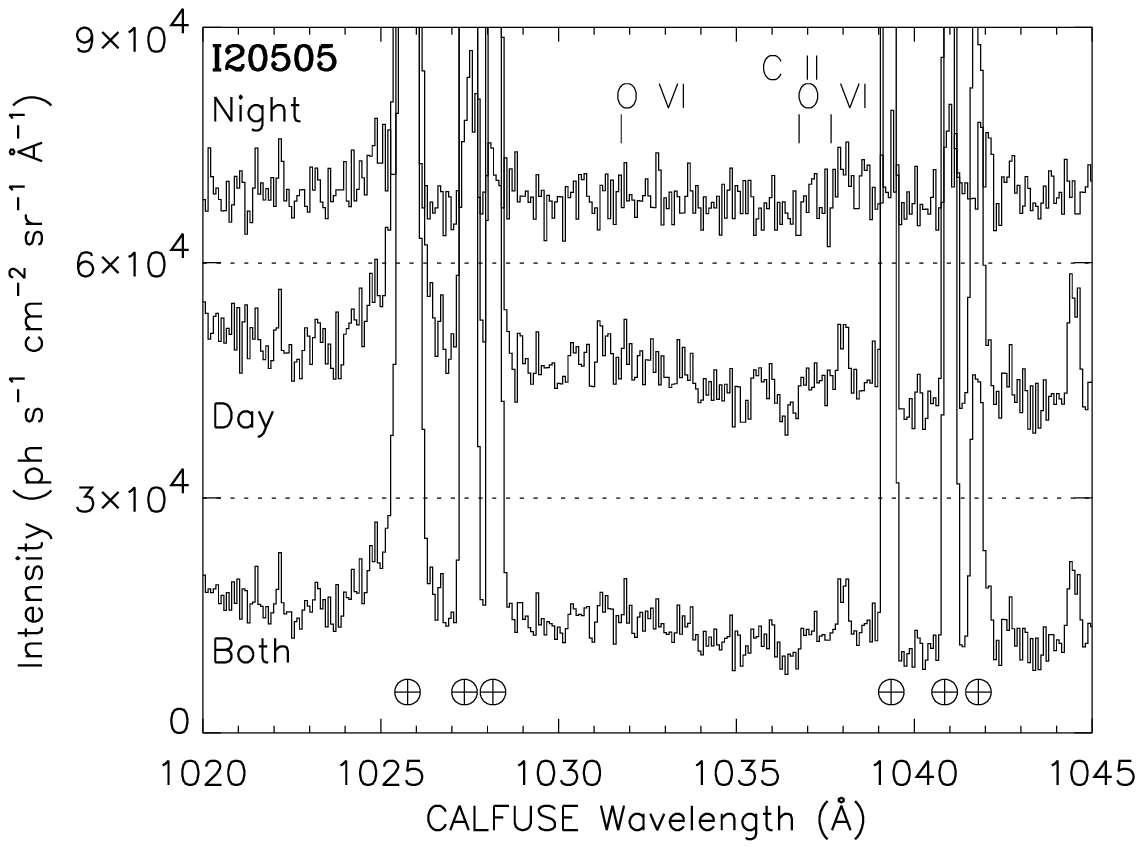}
\plotone{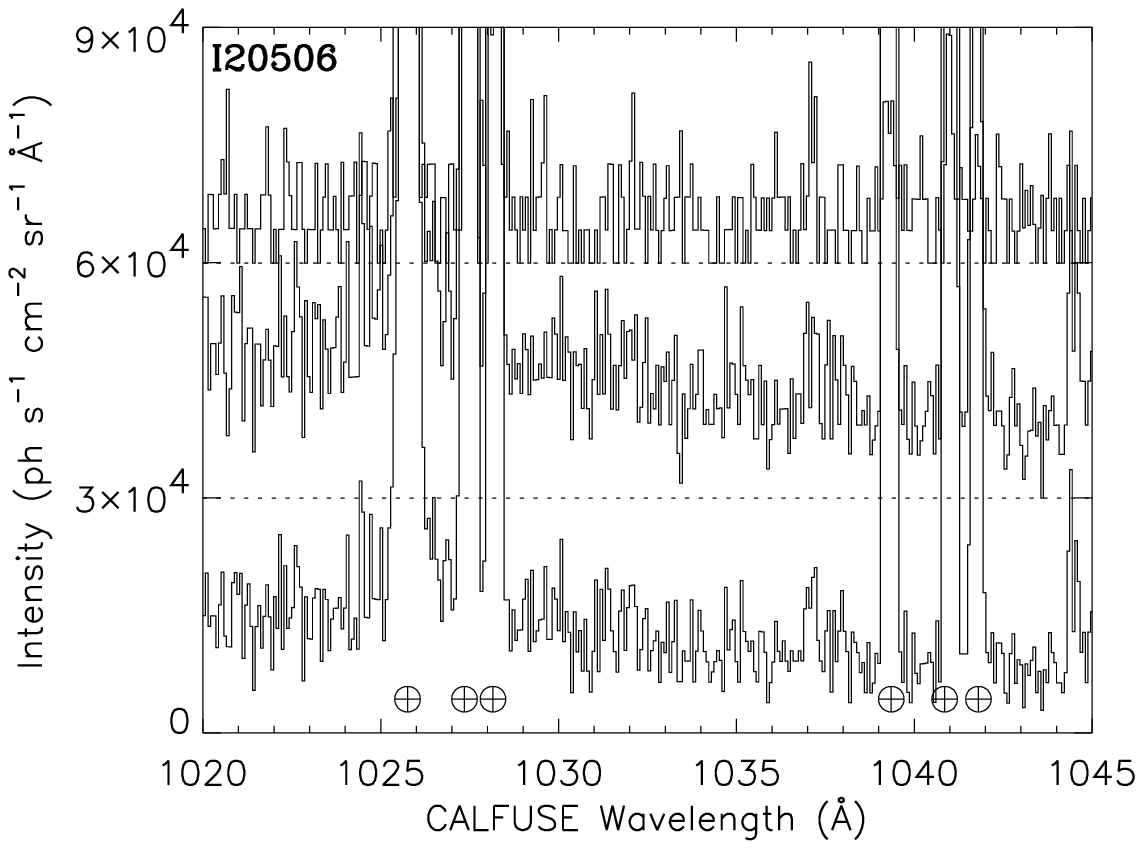}
  \caption{The 1020 to 1045 \AA\ region of the LiF 1A spectra extracted
from the B12901 (top panel), I20505 (middle panel) and
I20506 (bottom panel) data.  
Each panel presents the spectra drawn from the satellite-night
portion of the data (raised by
$6 \times 10^4$~photons~s$^{-1}$~cm$^{-2}$~sr$^{-1}$~\AA$^{-1}$),
the spectrum drawn from the satellite-day portion of the data (raised by
$3 \times 10^4$~photons~s$^{-1}$~cm$^{-2}$~sr$^{-1}$~\AA$^{-1}$),
and the spectrum drawn from the combined day and night data 
(unadjusted).  The wavelength scales have not yet been corrected.
Each spectrum has been binned by 11 pixels 
The Earth's atmospheric airglow lines are marked with 
$\oplus$ symbols.  The wavelengths of the 
\oxysix\ resonance lines and \cartwo\ 3s$^2$ S$_{1/2}$ to 2p$^2$ P$_{3/2}$
line are marked on the top and middle panels.  
No \oxysix\ or \cartwo\ emission is evident.}
\label{fig:lif1a}
\end{figure}

Figure~\ref{fig:lif1a} displays the 
1020 to 1045 \AA\ region of the LiF 1A spectra for the
three pointing directions.
At this stage, the {\it{absolute}} wavelength scales are still inaccurate,
though the {\it{relative}} wavelength scales of the pipeline spectra are
accurate to several detector pixels
(about 0.035 \AA\ in the LiF~1A spectrum).
The largest sources
of systematic error in the following intensity calculations
are the
uncertainties in the solid angle of the aperture and 
effective area calibration.  Both the LWRS's solid angle (30 arcsec $\times$ 
30 arcsec) and the LiF~1A effective area calibration
are thought to be accurately known to $\sim10\%$.  Thus,
the systematic uncertainty in intensity measurements 
will be $\sim14\%$.

\section{Spectral Analysis}

\begin{figure}
\epsscale{0.7}
\plotone{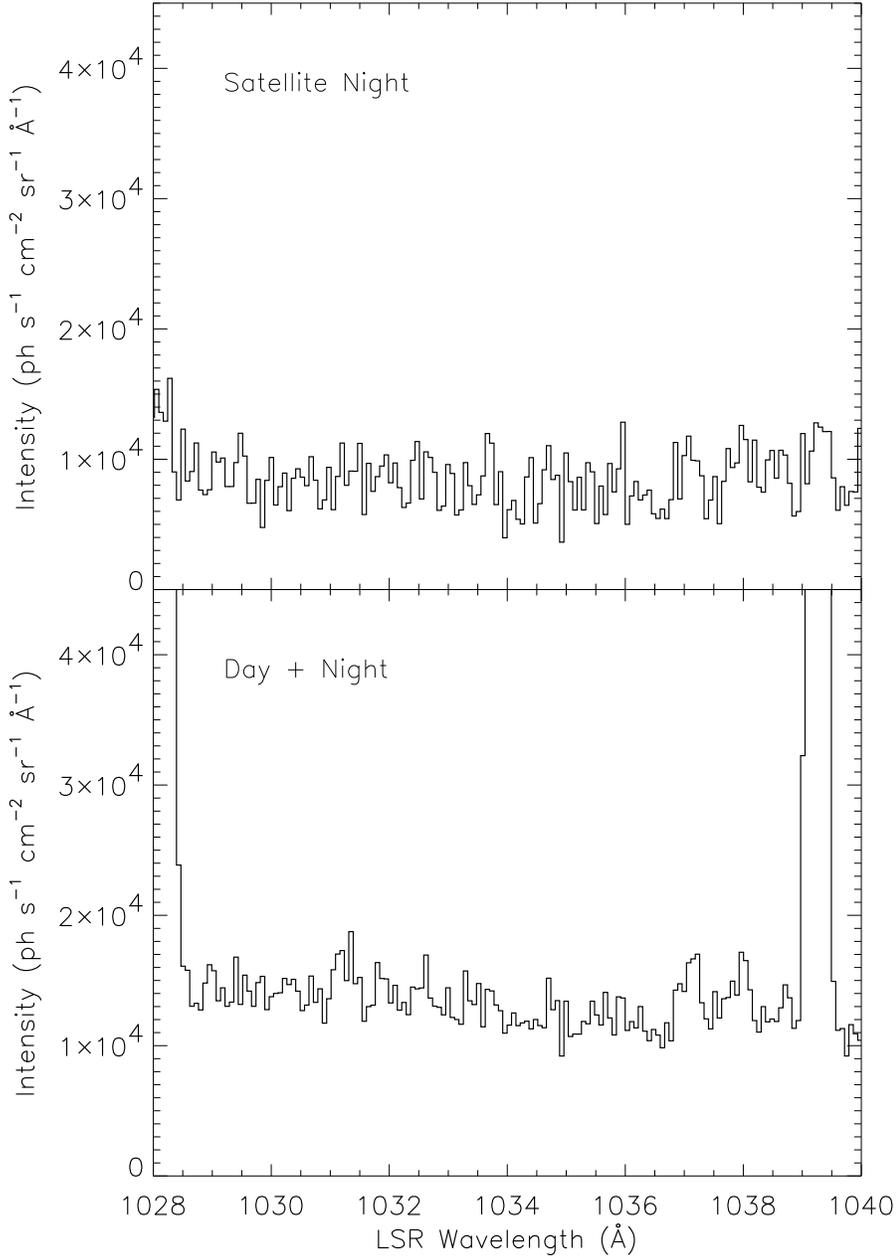}
\vspace{0.5cm}
  \caption{
Co-added spectrum from the night-only portion of the data
(top panel)
and co-added spectrum from both day and night portions of the data
(bottom panel).  The wavelength scale has been corrected and 
shifted to the LSR reference frame.  The spectral data has been binned by
11 pixels.  The night spectrum is flat, except for
variations due to noise.  No excess intensity
around 1032, 1038, or 1037 \AA\ is observed.  
Neither is there a 1032 \AA\ feature in the 
``Day$+$Night'' spectrum.
The sharp thin emission feature near 1031.5 \AA\
is only present during the 
satellite-day portion of the data and so is not attributable
to interstellar material.  Similar features have
appeared in previous daytime blank sky observations
\citep{shelton_etal}.
The ``Day$+$Night'' spectrum wavers significantly between
1036 and 1038 \AA, 
preventing us from searching for
the 1037 and 1038 \AA\ lines.}
\label{fig:coadd}
\end{figure}

In order to determine the absolute wavelength scales for the
LiF 1A spectra from the three pointings,
we used the airglow emission lines 
as reference wavelengths.  We then co-added the spectra 
from the three pointings and shifted
to the Local Standard of Rest (LSR) reference frame.
The 1020 to 1045 \AA\ region of the 
resulting satellite-night and day $+$ night spectra 
are displayed in Figure~\ref{fig:coadd}.

We found the continuum level in the satellite-night
spectrum by fitting a smooth curve to the 
surrounding spectral region.
We then searched the residual spectrum for emission
features.  None were found
in close proximity to the \oxysix\ resonance line rest wavelengths in the
LSR reference frame.  If cosmic features had existed,
their widths would have equaled or exceeded the instrumental
width function, their signals would have been calculated
from the numbers of counts in excess of the continuum
within the extraction width, and their 
random uncertainties would have been
calculated from the square roots of the numbers of spectral
counts (not residual counts) within the extraction width.
We have followed this method, using the total instrumental width 
(generously set to 0.43 \AA)
as the extraction width,
to determine the upper limits.  The ``signal'' and 1 sigma 
statistical uncertainty intensities
for the \oxysix\ resonance lines at their rest wavelengths in the
LSR reference frame are
180 $\pm$ 330 and 
-180 $\pm$ 310 photons cm$^{-2}$ s$^{-1}$ sr$^{-1}$, 
for the 1032 and 1038 \AA\ lines, respectively.  
These measurements are subject to the
above mentioned systematic uncertainties of $14\%$.
We validated our technique by applying the following alternate extraction
technique to sample spectra.  We fit the spectrum to a
continuum and small emission feature, then varied the size
of the feature to determine the 1 sigma uncertainties.  The
results were similar to those found by applying our techniques to the
same spectra.

When the satellite-day 
portions of the data sets were added, the spectrum became
more ragged.  From the variations between the
I20505, I20506, and B12901
satellite-day spectra
we surmise that the raggedness is primarily due to scattered light.
For the combined day and night spectrum, the ``signal'' and 1 sigma
statistical uncertainty intensities are
20 $\pm$ 230 and 80 $\pm$ 210 
photons cm$^{-2}$ s$^{-1}$ sr$^{-1}$, for the
1032 and 1038 \AA\ lines, respectively.  The measurements are
subject to the 
$14\%$ systematic uncertainties.
For reference, Table~\ref{table:measurements} tabulates these
results and lists the tightest 1 and 2 sigma upper limits.  
The 1 and 2 sigma upper limits were calculated from the unrounded
numbers.

We also searched for 
\carone, \cartwo, \carthree, \nitone, \nittwo, \nitthree, \magtwo, \sitwo, 
\stwo, \sthree, \sfour, \ssix, \irontwo, and \ironthree\
emission lines in the spectra taken with the LiF 1A, SiC 1B, LiF 2A,
and SiC 2A detector segments.  
We relied on the satellite-night portion of the data
when we searched for emission lines which also appear strongly in
the solar spectrum (i.e. \carthree, \citet{curdt_etal}),
appear weakly in the terrestrial airglow spectrum
(i.e. \nittwo, \citet{feldman_etal}), 
or are significantly stronger in the satellite-day portion
of the data than in the satellite-night portion of the
data (i.e. \cartwo).
The only signals detected were marginally significant
intensities of \carthree\ and
\nittwo\ (see Table~\ref{table:otherlines}).


\section{Discussion}

These observations are the first to place tight upper
limits on the intensity of the \oxysix\ emission from the
Local Bubble
(2 sigma upper limits on the 1032 and 1038 \AA\ emission lines of 
530 and 500 photons cm$^{-2}$ s$^{-1}$ sr$^{-1}$, 
respectively).  The theoretical ratio of the intensity of
1032 to 1038 \AA\ photons is 2.0 to 1.0.  
Using the 2.0 to 1.0 ratio and
the 1032 \AA\ line upper limit yields a
2 sigma doublet upper limit of 
800 photons cm$^{-2}$ s$^{-1}$ sr$^{-1}$ or
$1.5 \times 10^{-8}$ ergs cm$^{-2}$ s$^{-1}$ sr$^{-1}$.

\subsection{Comparisons with Local Bubble Models}


\noindent
{\it{Breakout Model:  $4 \times 10^4$~K, Overionized Bubble}}

In this scenario, a
series of supernova explosions and winds has created a hot, youthful
bubble.  It is assumed that the bubble expanded rapidly as it
broke into a low density region.
The rapid, adiabatic expansion cooled the bubble faster
than the ions could recombine, so that the
gas now has a temperture of 4 to $6 \times 10^4$~K,
but still highly ionized and X-ray emissive 
\citep{breitschwerdt_schmutzler, breitschwerdt}.  
Because the transition zones between the bubble plasma and the embedded
cool clouds will be cooler than the collisional
ionizational equilibrium temperature for \oxysix,
they need not be rich in \oxysix\ ions.

In the \citet{breitschwerdt} model,
the Local Bubble has a radius of 115 pc, density (of electrons,
Breitschwerdt, personal communication)
of $2.4 \times 10^{-2}$ cm$^{-3}$, and pressure (presumably thermal) of
$\sim 2000$~K~cm$^{-3}$.  
Thus, the plasma temperature is approximately $4.3 \times 10^4$~K.
Lacking published estimates for the \oxysix\ intensity, we will
calculate it from the above values
and the emission equation in \citet{shull_slavin}.
The intensity calculation also requires 
the ratio of electrons to hydrogen nuclei in a solar abundance
plasma (1.2), the solar abundance of oxygen
($8.5 \times 10^{-4}$, \citet{grevesse_anders}), and
an estimate of the 
fraction of oxygen atoms in the \oxysix\ ionization level.
During the plasma's extraordinary cooling history, it
has experienced some degree of recombination, as evidenced
by the strong lines of \carthree, \oxythree, and \oxyfour\
in \citet{breitschwerdt}'s predictions.  Here,
we take the lower limit on recombinations of oxygen to
the \oxysix\ state as that of a
$5 \times 10^5$~K collisional 
ionizational equilibrium plasma.  
In this case, the fraction of 
oxygen in the O VI state is 0.037 
\citep{mazzotta_etal,schmutzler_tscharnuter}.
As the plasma recombines further, the \oxysix/oxygen fraction increases, and,
due to recombinations of \oxyseven,
does not decrease until the most of the oxygen has recombined
to \oxyone\ and \oxytwo.
An upper limit on the extent of the recombinations
is taken from the isochoric cooling predictions of 
\citet{schmutzler_tscharnuter} at the model temperature,
again yielding an \oxysix\ to oxygen fraction of 0.037.
We take 0.037 as our estimate, noting that the \oxysix/oxygen ratio and the
intensity prediction may be higher.
The resulting intensity estimate,
$\sim1900$ photons cm$^{-2}$ s$^{-1}$ sr$^{-1}$,
is more than twice the 2 sigma upper limit established in this paper.
%

For this calculation, the \oxysix\ ions are assumed to be 
mixed throughout the remnant. 
If the assumed abundance of metal atoms were reduced from
the solar value, then the soft X-ray scaling relationships
used in finding the model parameters
would need to be adjusted to a similar degree.   As a result,
the model parameters would need to be adjusted, probably leading to
a bigger, denser, and/or hotter bubble.  
As far as the predicted \oxysix\ intensity is concerned,
these adjustments would probably offset the hypothetical 
abundance adjustment.

Not only is there a significant discrepancy 
between our null result and the intensity
calculated from the model, there is a factor of 20 difference
between the observed \oxysix\ 
column density ($\sim1.6 \times 10^{13}$ cm$^{-2}$, 
\citet{shelton_cox}, \citet{oegerle_etal},
see below) and the column density
calculated from the model ($2.7 \times 10^{14}$ cm$^{-2}$), and
a significant discrepancy between 
the 2 sigma upper limit on the observed \carthree\ 977 \AA\
intensity 
($7300$ photons cm$^{-2}$ s$^{-1}$ sr$^{-1}$)
and the published \carthree\ prediction 
($\geq 8500$ photons cm$^{-2}$ s$^{-1}$ sr$^{-1}$, 
Figure~2 \citet{breitschwerdt}).
The number and magnitude of these discrepancies 
practically eliminate this class of models.  

\noindent
{\it{Generalized Multi-Phase Model of the Local Bubble}}

The lesson learned from evaluating the previous model was that 
a bubble consisting of a single, moderate temperature, overionized plasma
is disallowed because it
contains too many \oxysix\ ions and emits too many photons
from its \oxysix\ and \carthree\ ions.
Furthermore, a bubble consisting of a single, 
collisional ionizational equilibrium plasma which is hot enough
to produce the observed soft X-ray surface brightness is disallowed
because it contains too few \oxysix\ ions (see below).  
In contrast, suppose the Local Bubble contains
multiple distinct plasmas,
a collisional ionizational equilibrium, 
million degree, X-ray emissive plasma and 
a several hundred thousand degree, \oxysix-rich plasma.
Cool clouds are also known to lie within the Local Bubble so
such a topology is reasonable.  One example of an allowed form is 
a hot bubble, bounded by transition temperature gas and
possibly containing additional transition zones around
embedded cool clouds.
Here, we consider such a generalized multi-phase model, in order to
determine if it can meet the observational constraints or if
more unusual circumstances are required.

The hotter component in this multi-phase model accounts for the observed
soft X-rays, but few \oxysix\ photons.
Based on the temperature found from a soft X-ray analysis
($10^{6.1}$~K, assuming collisional ionizational equilibrium,
\citet{kuntz_snowden}),
the observed {\it{ROSAT}} \ R12 
band countrate seen toward this part
of the filament (590 $\times 10^{-6}$ counts s$^{-1}$ arcmin$^{-2}$,
K. D. Kuntz, private communication), the
conversion between the {\it{ROSAT}} \ R12 band countrate and
the emission measure ($1 \times 10^{-6}$ counts s$^{-1}$ arcmin$^{-2}$
$=$ $7.15 \times 10^{-6}$ cm$^{-6}$ pc$^{-1}$, \citet{snowden_etal_97}), 
the fraction of oxygen atoms in the \oxysix\ ionization level at 
this temperature
($3.2 \times 10^{-3}$, \citet{mazzotta_etal}),
the solar abundance of oxygen 
($8.5 \times 10^{-4}$, \citet{grevesse_anders}),
the ratio of free electrons to hydrogen nuclei (1.2),
and the electron-impact excitation rate coefficient of \citet{shull_slavin},
we estimate that the $\sim10^6$~K region 
produces only
$\sim$60 photons cm$^{-2}$ s$^{-1}$ sr$^{-1}$ in the \oxysix\ resonance
line doublet from a column density of
only $5 \times 10^{11}$ \oxysix\ ions cm$^{-2}$.

The primary source of \oxysix\ ions and resonance line photons would
be the transition temperature gas.  
Althought it need not be in collisional ionizational
equilibrium, we will assume its temperature is 
between $1 \times 10^5$~K and $1 \times 10^6$~K for the following calculations.
The Local Bubble's column density of \oxysix\ ions is
roughly $1.6 \times 10^{13}$ cm$^{-2}$ (see following subsection).
By assuming that the gas is in thermal pressure balance with 
the soft X-ray emitting plasma
($P_{th}/k \sim$ 15,000 K cm$^{-3}$, \citet{snowden_etal_98}), 
we estimate its electron density.
The temperature is not well known, but
most of the \oxysix\ ions are likely to be 
near their ionizational equilibrium 
temperature ($3.2 \times 10^5$~K), and
the emission equation 
is relatively insensitive to temperature for 
$1 \times 10^5 \stackrel{<}{\sim} 
T \stackrel{<}{\sim} 1 \times 10^6$~K.
An \oxysix-rich plasma, with the above column density,
a temperature of $6.3 \times 10^5$~K, and a thermal pressure
of 15,000 K cm$^{-3}$ emits 
450 photons cm$^{-2}$ s$^{-1}$ sr$^{-1}$, 
which is within the observational 2 sigma upper limit on the intensity.
Similar plasmas with temperatures as low as 
$\sim 1 \times 10^5$~K emit even less intense \oxysix\ radiation.
Thus, a two phase model could meet the observational constraints without
requiring unobserved astrophysics or unusual conditions. 


\noindent
{\it{Simulations of A Hot Bubble with Transition Zones}}

In this section, we examine the multi-phase bubble simulations
of \citet{smith_cox} and the evaporating cloud simulations of
\citet{slavin}.
In the models of \citet{smith_cox},
the Local Bubble resulted from two or three supernova
explosions occurring up to several million years ago.
The interior of the structure
consists of hot ($\sim 10^6$~K),
nearly collisional ionizational equilibrium plasma, while
the periphery of the structure consists of 
an intermediate temperature transition zone.
Although the simulations do not explicitly include the
transition zones surrounding the embedded cool clouds, 
such zones are thought to 
harbor \oxysix\ ions \citep{slavin,oegerle_etal} 
and so should contribute to the
flux of \oxysix\ resonance line photons.  Here, we will use the
simulations of \citet{slavin}, which predict their intensity.

After performing several hydrodynamical simulations, 
\citet{smith_cox} 
decided that the best model lies intermediate between
their various simulations.
Their simulations yield predicted \oxysix\ intensities ranging from 
190 to more than 9,500 photons cm$^{-2}$ s$^{-1}$ sr$^{-1}$,
(after the extra factor of $4\pi$ in their
Figure 19 (Smith, private communication) is divided out).
To this intensity must be added the intensity emitted by the
transition zone on the local cloud and, possibly, transition zones
on other cool clouds embedded in the Local Bubble.
For the transition zone on the cool cloud surrounding the Sun,
\citet{slavin} predicted an \oxysix\ doublet
intensity of 250 photons cm$^{-2}$ s$^{-1}$ sr$^{-1}$.  
The sum of the \citet{smith_cox} and \citet{slavin} predictions 
(440 to $\geq9800$~photons cm$^{-2}$ s$^{-1}$ sr$^{-1}$) 
marginally overlaps the
2 sigma upper limit reported in this paper.
If other cool clouds reside along the line of sight, then the
predicted \oxysix\ intensity would be even higher.
Thus, this particular model is heavily constrained by the null results.
We would have lost confidence in this type of model had it not
been possible to simultaneously satisfy the 
\oxysix\ emission, \oxysix\ column density, and
{\it{ROSAT}} soft X-ray constraints with the above
generalized multiphase model.

\subsection{Limits on Electron Density, Pressure, Pathlength, Cooling
Timescale}

Here, we will combine the intensity upper limit
with column density estimates taken from the literature 
in order to calculate the upper limits on the 
electron density and thermal pressure in the \oxysix-rich plasma.
The first data-derived estimate of the 
typical \oxysix\ column density of the Local Bubble 
($N_{OVI} \sim 1.6 \times 10^{13}$ cm$^{-2}$, \citep{shelton_cox})
was found via a statistical analysis of 
several dozen {\it{Copernicus}} lines of sight, most of which 
terminated far beyond the Local Bubble.
Later, \fuse\ observed a couple dozen lines of sight 
entirely within the Local Bubble.
The work of \citet{welsh_etal} and \citet{oegerle_etal} 
suggest that the \oxysix\ column density {\it{within}} the Local Bubble
may be as little as a few $\times 10^{12}$ cm$^{-2}$ in some directions,
but that the column density increases roughly with distance from
the Sun and the total column density along a 100 pc radial path
is consistent with values of $7 \times 10^{12}$ cm$^{-2}$
to $1.6 \times 10^{13}$ cm$^{-2}$.
Here, we take the column density from the published survey
(i.e. $1.6 \times 10^{13}$ cm$^{-2}$).

Using the
emission equation of \citet{shull_slavin},
the above column density,
the 2 sigma upper limit on the doublet intensity,
and the temperature
at which \oxysix\ is most prevalent in collisional ionizational
equilibrium plasmas
($T = 3.2 \times 10^5$~K, \citet{mazzotta_etal})
yields an estimated upper limit for the electron density,
$n_e$, of 0.021 cm$^{-3}$.  
Thus, the estimated upper limit for 
the thermal pressure, from $P_{th}/k = 1.92 n_e T$, 
is $13,000$~K~cm$^{-3}$.  
If the temperature were significantly higher than the assumed temperature,
then the pressure would be greater.
If it is not, then
the total (thermal and nonthermal) pressure in the \oxysix-rich 
region of the Local Bubble (presumably the transition zones)
could be brought into balance with that of the X-ray
emitting region of the Local Bubble (presumably the interior, 
$P_{th}/k \sim$ 15,000 K cm$^{-3}$)
by assuming that the cooler, denser \oxysix-rich regions 
have a greater non-thermal pressure than the hotter, more 
rarefied X-ray emitting region.

Furthermore, 
the \oxysix-rich material occupies a pathlength ($\Delta l$) of 
at least ${\sim}5900 N_{OVI}/n_e$ in cm, (incorrectly
type-set in \citet{shelton}),
or $\stackrel{>}{\sim} 0.15$~pc, consistent with transition zones.
The cooling timescale solely due to cooling through the \oxysix\ doublet is 
\begin{equation}
t = \frac{\frac{3}{2} k T}{4 \pi {\rm{(sr)}} I_{OVI}}
N_{OVI} 
\frac{n_T}{n_{ovi}} ,
\end{equation}
where $k$ is Boltzman's constant, $n_{ovi}$, $n_{oxy}$, 
${n_H}$, and ${n_T}$
are the densities of \oxysix\ ions, oxygen atoms, hydrogen
nuclei, and particles, respectively, sr means steradians,
and $n_{ovi}/n_{oxy} = 0.24$ at $3.2 \times 10^5$~K
\citep{mazzotta_etal}.
Thus, the cooling timescale due solely to \oxysix\ resonance
line emission is at least $2.0 \times 10^6$~years.
These Local Bubble values
are tabulated in Table~\ref{table:calc}, along with the
Galactic halo values calculated in the following subsection.

\subsection{Galactic Halo}

Previous \fuse\ studies of emission from \oxysix\ in the interstellar
medium (ISM) outside of superbubbles or supernova remnants
observed far longer paths
(\citet{shelton_etal,dixon_etal,shelton}; and \citet{welsh_etal}).
In each of those cases, the instrument had been directed toward a high
Galactic latitude direction ($|b| > 40^{\rm{o}}$).  Given that the
local region contributes little
\oxysix\ resonance line intensity,  the vast majority of the
intensity observed on the relatively unobscured, high latitude observations
must originate beyond the Local Bubble.
Extragalactic sources are ruled
out by the velocities of the observed emission features and
so this emission presumably resides in the Galactic thick disk and halo.
Subtracting the Local Bubble's doublet intensity
from the averages of the intensities observed by
\citet{shelton_etal,dixon_etal,shelton}; and \citet{welsh_etal}
leaves 
3500 to 4300 photons cm$^{-2}$ s$^{-1}$ sr$^{-1}$,
attributable to the Galactic halo.

The column density through the halo is roughly 
$3.0 \times 10^{14}$ cm$^{-2}$ \citep{savage_etal}.
Thus, using the method of Section 5.2,
we find the electron density in the halo's \oxysix-rich
gas to be 0.0050 to 0.0061 cm$^{-3}$.
The thermal pressure is 3100 to 3800 K cm$^{-3}$,
consistent with the range of {\bf{total}} pressures 
in models of the
Galactic thick disk and lower halo \citep{boulares_cox,ferriere}.
The pathlength is at least 110 pc.
The cooling timescale is 7 to $8 \times 10^6$ years.


\section{Summary}

\begin{itemize}

\item 
The Local Bubble's \oxysix\ intensity has been examined
through a shadowing observation directed toward
$l = 278.6^{\rm{o}},b = -45.3^{\rm{o}}$.  No emission
was seen.  The tightest 2 sigma upper limits on the 
1032 and 1038 \AA\
emission lines are 530 and 500 photons cm$^{-2}$ s$^{-1}$ sr$^{-1}$,
respectively.  Given the theoretical intensity
ratio (2.0 to 1.0), the tightest 2 sigma upper limit on
the resonance line doublet is 800 photons cm$^{-2}$ s$^{-1}$ sr$^{-1}$.

\item 
The class of models in which the Local Bubble is assumed to
be a single, moderate temperature ($\sim4 \times 10^4$~K), extremely
overionized plasma is virtually
ruled out.  The 2 sigma upper limits on the \oxysix\ and \carthree\
intensities exceed the model predictions, as does the
observed \oxysix\ column density.
The class of models in which the Local Bubble
is a single, hot ($10^6$~K), collisional ionizational equilibrium plasma
are also ruled out.  Such a model could not produce the observed
\oxysix\ column density.
However, the combined \oxysix\ and {\it{ROSAT}}
soft X-ray observations allow models in which the Local Bubble
contains multiple, separate phases.  The X-ray emission can be
attributed to the hot zone and the \oxysix\ ions can be attributed
to transition zones.
Existing simulations of a multiphase bubble combined with
simulations of a single evaporating cloud are tightly constrained
by our 2 $\sigma$ upper limit on the doublet intensity.  Considering that
multiple clouds are expected to reside along the typical line of
sight, the expected \oxysix\ intensity falls well below the
upper limit, suggesting that transition zones differ from current
models.

\item 
The estimated 2 sigma upper limit on the
electron density in the Local Bubble's \oxysix-bearing plasma is
0.021 cm$^{-3}$.
If the temperature is near the collisional ionizational
equilibrium temperature for \oxysix\ ($3.2 \times 10^5$~K),
then the thermal pressure is $\leq13,000$~K cm$^{-3}$, which is
weaker than that of the soft X-ray emissive portion of the
Local Bubble.  The discrepancy could be resolved if 
the \oxysix-bearing gas were to
have a greater than assumed temperature or a significant
nonthermal pressure.

\item 
The timescale for cooling soley due to the \oxysix\ resonance
line emission is proportional to the thermal energy in
the gas and inversely proportional to the \oxysix\ intensity.
For the Local Bubble, this timescale is 
$\geq 2 \times 10^6$ years.  

\item 
The Galactic halo's \oxysix\ doublet intensity has
been found by subtracting the Local Bubble result
from other observations.  It is 
3500 to 4300 photons cm$^{-2}$ s$^{-1}$ sr$^{-1}$.

\item 
The Galactic halo's intensity implies an electron
density of $\sim0.0050$ to $\sim0.0061$ cm$^{-3}$.
If the temperature is near the collisional ionizational
equilibrium temperature for \oxysix\,
then the thermal pressure is
$\sim3100$ to $\sim3800$ K cm$^{-3}$.
This pressure is consistent with the total pressure
estimates in models of the thick disk and lower
halo.

\item 
The \oxysix\ in the Galactic halo cools on 
a timescale ($t \propto$ thermal energy / \oxysix\ intensity)
of 7 to $8 \times 10^6$ years. 

\end{itemize}

\acknowledgements

We thank 
the referee, Steve Snowden, for volunteering the {\it{DIRAS}} corrected
{\it{IRAS}} and
{\it{ROSAT}} images and critiquing the manuscript,
K. D. Kuntz for providing the ROSAT measurements and
commenting on the manuscript,
B.-G. Andersson for discussing the telescope's specifications,
and Tim Heckman for reviewing the manuscript. 
We acknowledge \citet{welsh_etal}, which compared observed and
modeled \carthree\ intensities for other directions.
This research has been supported 
by NASA grant NAG5-10394 (through the \fuse\ Guest Investigator program)
and NASA grant NAG5-10807 (through the Long Term Space Astrophysics
Program).

\pagebreak

\clearpage

\begin{deluxetable}{lcccc}
\tablewidth{0pt}
\tablecaption{}
\tablehead{
\colhead{Program Id }     
& \colhead{Start Date}
& \colhead{LiF 1A Exposure Time}
& \colhead{Night Time}
& \colhead{LWRS Direction}\\
{Number}
& {(UT)}
& {(ksec)}
& {(ksec)}
& {(l,b)}}
\startdata
I2050501 & Aug. 21, 1999 & 84.5  & 27.8 & $278.58^{\rm{o}},-45.31^{\rm{o}}$\\ 
I2050510 & Aug. 24, 1999 & 17.7  & 4.5  & ''\\
I2050601 & Aug. 27, 1999 & 18.2  & 5.6  & $278.59^{\rm{o}},-45.30^{\rm{o}}$\\
B1290101 & Aug. 13, 2001 & 52.5  & 9.6  & $278.63^{\rm{o}},-45.31^{\rm{o}}$\\
B1290102 & Aug. 16, 2001 & 55.0  & 15.4 & ''\\ \hline
Total    &               & 227.7 & 62.8 & $278.6^{\rm{o}},-45.3^{\rm{o}}$ \\
\enddata
\label{table:data}
\end{deluxetable}

\begin{deluxetable}{lcc}
\tablewidth{0pt}
\tablecaption{Upper Limits on the \oxysix\ Emission Lines}
\tablehead{
\colhead{Emission Region}
& \colhead{Intensity (ph s$^{-1}$ cm$^{-2}$ sr$^{-1}$)}     
& \colhead{Intensity (ph s$^{-1}$ cm$^{-2}$ sr$^{-1}$)} \\
\colhead{}
& \colhead{Night Only}     
& \colhead{Day$+$Night}}
\startdata
	       & Combined Dataset &	\\ \hline
1031.93 \AA\  & $180 \pm 330 \pm 14\%$ & $20 \pm 230 \pm 14\%$  \\ 
1037.62 \AA\  & $-180 \pm 310 \pm 14\%$ & $80 \pm 210 \pm 14\%$ \\ \hline
	       & Tightest 1 Sigma Limits, incl. $14\%$ Systematic Uncertainty &	\\ \hline
1031.93 \AA\  &                         &  280  \\ 
1037.62 \AA\  & 150			&           \\ \hline
	       & Tightest 2 sigma Limits, incl. $14\%$ Systematic Uncertainty &	\\ \hline
1031.93 \AA\  &                         &  530 \\ 
1037.62 \AA\  & 500			&           \\ 
\enddata
\label{table:measurements}
\end{deluxetable}
\pagebreak

\begin{deluxetable}{llll}
\tablewidth{0pt}
\tablecaption{Other Transitions}
\tablehead{
\colhead{Species}     
& \colhead{Rest Wavelength} 
& \colhead{Intensity} 
& \colhead{Notes}  \\
\colhead{}     
& \colhead{(\AA)} 
& \colhead{(photons cm$^{-2}$ s$^{-1}$ sr$^{-1}$)}
& \colhead{} 
}
\startdata
C~{\footnotesize{I}}    & 945.6 \AA\    & $-300 \pm 390$ & B \\
C~{\footnotesize{I}}    & 1122.3 \AA\   & $-240 \pm 370$ & N \\
C~{\footnotesize{II}}   & 1037.0 \AA\   & $30 \pm 290$ & N \\
C~{\footnotesize{III}}  & 977.0 \AA\    & $4700 \pm 1300$ & N, SiC 1B \\
                    ''  &         ''    & $2600 \pm 1000$ & N, SiC 2A \\
N~{\footnotesize{I}}    & 954.0 \AA\    & $-200 \pm 740$ & N\\
N~{\footnotesize{II}}   & 916.7 \AA\    & $2300 \pm 1100$ & N, SiC 1B \\
N~{\footnotesize{III}}  & 991.6 \AA\    & $-170 \pm 850$ & N\\
Mg~{\footnotesize{II}}  & 946.7 \AA\    & $-430 \pm 680$ & N\\
Si~{\footnotesize{II}}  & 1023.7 \AA\   & $-200 \pm 190$ & B\\
S~{\footnotesize{II}}   & 906.9 \AA\    & $-240 \pm 1100$ & N\\
S~{\footnotesize{III}}  & 1015.5 \AA\   & $170 \pm 240$ & B\\
S~{\footnotesize{IV}}   & 1062.7 \AA\   & $-390 \pm 300$ & N\\
S~{\footnotesize{IV}}   & 1073.0 \AA\   & $300 \pm 280$ & N\\
S~{\footnotesize{VI}}   & 933.4 \AA\    & $170 \pm 690$ & N\\
S~{\footnotesize{VI}}   & 944.5 \AA\    & $1000 \pm 840$ & N\\
Fe~{\footnotesize{II}}  & 1144.9 \AA\   & $270 \pm 220$ & B\\
Fe~{\footnotesize{III}} & 1122.5 \AA\   & $-700 \pm 350$ & N\\ \hline
\enddata
\begin{flushleft}
The reported error bars reflect the 1 sigma random uncertainties.
All measurements are subject to the net systematic uncertainty of $14\%$.
``B'' denotes both the day and night portions of the
data and ``N'' denotes the satellite-night portion of the
data.  In cases where airglow emission or scattered light affected
the satellite-day data for the spectral region of interest
or the satellite-day spectrum undulated substantially, we used only
the satellite-night portion of the data.  We also used the night
spectrum for the Mg~{\footnotesize{II}} observation 
because it led to a more constraining upper limit.  For the
measurements in this table, the tighter
pulseheight cuts were used in processing the B12901 portion of the
data.
\end{flushleft}
\label{table:otherlines}
\end{deluxetable}
\pagebreak

\begin{deluxetable}{lcccc}
\tablewidth{0pt}
\tablecaption{}
\tablehead{
\colhead{Region}     
& \colhead{$n_e$}
& \colhead{$P_{th}/k$}
& \colhead{$\Delta l$}
& \colhead{$t$}\\
{}
& {(cm$^{-3}$)}
& {(K~cm$^{-3}$)}
& {(pc)}
& {(yr)}}
\startdata
Local Bubble	& $\leq 0.021$ 
		& $\leq 1.3 \times 10^4$ 
		& $\stackrel{>}{\sim} 0.15$ 
		& $\geq 2 \times 10^6$ \\
Galactic Halo   & 0.0050 to 0.0061    
		& 3100 to 3800 
		& $\stackrel{>}{\sim} 110$ 
		& 7 to $8 \times 10^6$ \\
\enddata
\label{table:calc}
\end{deluxetable}
\pagebreak

\clearpage

\end{document}